\documentstyle[12pt]{article}
\input psfig
\long\def\comment#1{}
\begin{document}
\title{Are Quantum Computing Models Realistic?}
\author{Subhash Kak\thanks{Department of Electrical \& Computer Engineering,
Louisiana State University,
Baton Rouge, LA 70803-5901, USA; {\tt kak@@ece.lsu.edu}}}
\maketitle

\begin{abstract}
The commonly used circuit model of quantum computing 
leaves out the problems of
imprecision in the
initial state preparation, 
particle statistics (indistinguishability of particles
belonging to the same quantum state),
and error correction (current techniques cannot correct all
small errors).
The initial state in the circuit model computation is obtained
by applying potentially imprecise Hadamard gate operations
whereas useful quantum computation requires a state with no
uncertainty.
We review some limitations of the circuit model and speculate on
the question if a hierarchy of 
quantum-type computing models exists.
\end{abstract}


\subsection*{Introduction}

Quantum information science provides important insights into several
aspects of the communication and computing process.
Single qubits have been proposed for use in
cryptography.
Pairs of entangled qubits can, in principle, have
remarkable applications:
two bits of classical information may be exchanged
using an existing entangled pair with the
two parties while transferring only one qubit by means of
the protocol of dense coding; and
an unknown quantum state may be teleported
to another location by use of an entangled pair of
qubits and classical bits so long
as the entangled qubits do not have any
phase uncertainty associated between them [1].

When we go from single and entangled
pairs of particles to groups 
of particles, as in various methods
of quantum computing, the question of the
physical realizability of the mathematical 
model become more problematic.
For example, the circuit model of quantum computing [2] 
leaves out problems of state preparation (how
to get the individual particles into a precise
state using gates that would have imprecision associated
with them),
particle statistics (indistinguishability of quantum particles
of the same quantum state),
and effective error correction.
It is assumed that once the qubits, each placed into a
superposition of $|0\rangle$ and $|1\rangle$ by the use
of an appropriate rotation operator, are loaded individually
on the $n$-cell register, Hamiltonians for the subsequent
evolution of the set of $n$-qubits will somehow be
found.
The physical implementability of the unitary matrices
is not addressed.
The model of computing also does not address the questions
of statistics and error correction.

The quantum Turing machine and the quantum cellular automata
models are equivalent to the circuit model and, therefore,
face the same difficulties.
These models, inspired by the philosophically extravagant
many worlds interpretation of quantum mechanics [3],
assign specific information to the qubits, postulating
gates that implement the unitary 
transformation representing the solution to the computational
problem.

The quantum circuit model converts the physical problem to a circuit-theoretic
form but it does not map all the physical constraints required
by the laws of quantum mechanics.
It gives specific labels to different
lines of the circuit and does not consider the question of
the indistinguishability of
particles in quantum mechanics.
This indistinguishability 
may require constraints additional to the
ones that are usually assumed when considering implementation.

It is good to remember that ``a {\it quantum system} is 
a useful abstraction, which frequently appears in the 
literature, but does not really exist in nature'' [4].
Quantum computing models use selected elements
of this abstraction in a manner that may preclude successful
physical implementation.
If a quantum computing model is not physically
implementable, then it should be called a
quasi-quantum model.

The quantum computing model -- like the billiard ball model [5] --
is an example of
a Hamiltonian system. 
Several years ago, Rolf Landauer cautioned [6] against
the Hamiltonian approach to computation.
In contrast to digital computers where 
data is reset, a Hamiltonian system cannot
correct local errors.
Quantum error correcting
codes have been proposed but they can only
correct certain large errors without correcting
small errors.
Even in theory these codes work only to correct bit-flips
and phase-flips, which is a vanishing small fraction of
all the phase errors that can occur in the quantum state.
Besides successful error correction, coding requires
that the error be within bounds, whereas the
uncertainty with regard to phase makes that 
assumption invalid.

The question of decoherence of quantum
states is another problem afflicting
quantum computation
but it does not concern us here.

There is also the question of the fundamental
limitations of the quantum computing paradigm.
Its unitary evolution is unable to perform basic
nonlinear mappings.
For an unknown state $x$, a general unitary matrix
$U$ does not exist which will take $|x0\rangle$
to $|xx\rangle$ or vice versa. In other words,
an unknown state can neither be copied nor deleted.
These operations are nonlinear and they are
beyond the capacity of a
unitary transformation.

By carrying the input data alongside, one can convert [7]
a one-way mapping to a reversible mapping, but that
would involve an exponential growth of overhead in any
substantial computation and, therefore, this possibility
cannot be taken seriously for real computational tasks.
As unitary transformations, quantum algorithms would
still be useful in certain problems, but this usefulness
would be similar to that of optical computing.
Since unitary mappings are rotations on a sphere (of
high dimensionality), one can only hope to compute periodicity
information or properties that can be related to this information.

In this note I list some
interrelated issues related to the quantum circuit model.
I first review the problems of creating an appropriate
pure state to get the computation started and then
consider the question of quantum statistics in the context
of such a state.
The thesis of this note is that ``quantum computing'' models
use the mathematical apparatus of quantum theory but do not
appear to incorporate all of its restrictions.
If this thesis is correct then one
may ask if other 
mathematical models of distinct computing power exist.

\subsection*{On the realizability of the circuit model}
The circuit model of quantum computing provides 
a schematic realization
of the unitary matrix that represents the computation in terms
of its submatrices.
It is implicit that when such transformations
are applied to the qubits on the register the evolution
will correspond to the quantum evolution given by
the Schr\"{o}dinger equation. 
This is correct but for the fact that the
circuit model takes the qubits to be unique and
distinguishable from each other, 
a condition that maps into the uniqueness of the
wires in the quantum circuit.
But quantum
objects cannot be distinguished amongst each other
before measurement. 
From a practical point of view it imposes severe constraints
on the labels that are ascribed to qubits.
This could mean that the unitary matrices for certain
gates may not be physically realizable.
The circuit model may then be seen as
an implementation not of quantum physics but of 
unitary transformations.

In the circuit model the
register is loaded with data one qubit at a time
where these qubits are independent of each other.
Now Hadamard transformation is applied to each qubit.
From a practical point of
view, due to the imprecision in the implementation of
the transformations, 
this will create a compound pure state with uncertain
weights.

In several proposed implementations, the individual qubits
themselves are not in a pure state. 
One must remember that a pure state must yield a predictable
outcome in {\it a specified maximal test} [8], and no such test
may be conceptualized for the qubits on the quantum register
in certain practical systems [9].

\subsection*{Unknown phase}
The state function of a quantum system is defined on
the complex plane whereas observations can
only be real. This means that the state function
may not be completely known even if the state is
prepared because of the uncertainty associated
with the state preparation process itself.
In such a situation one cannot hope to characterize
this reality with such precision so as to carry out a specific
computation using a single quantum state.

In general there may be unknowable phase associated
with the qubits [10] making it impossible to rotate
this qubit through a precise angle [11].
For convenience assume that the operator

\begin{equation}
M = \frac{1}{\sqrt 2} \left[ \begin{array}{cc}
                                  1 & 1 \\
                                  1 & -1 \\
                               \end{array} \right]
\end{equation}
is implementable.
When applied to the qubit $  \frac{1}{\sqrt 2}( |0\rangle +  |1\rangle),$
it will lead to the pure state $|0\rangle$.
But since the qubit should be realistically seen to be
$\frac{1}{\sqrt 2} (e^{i \theta_1}  | 0\rangle + e^{i \theta_2} | 1\rangle )$
(because of the imprecision in the gate),
an operation by M will take the qubit only to

\begin{equation}
\frac{(e^{i \theta_1} + e^{i \theta_2})}{2} | 0\rangle  +
\frac{(e^{i \theta_1} - e^{i \theta_2})}{2} | 1\rangle
\end{equation}

The probability of obtaining a $|0\rangle$ will now
be
$ \frac{1}{2} [1 + cos ( \theta_1 - \theta_2)],$
whereas the probability of obtaining a $ | 1\rangle$ will be
$ \frac{1}{2} [1 - cos ( \theta_1 - \theta_2)].$
The probabilities for the basis observables are
not exactly $\frac{1}{2}$, and they depend on the
starting {\em unknown} $\theta$ values.
Thus, the qubit can end up anywhere on the unit circle.
As example, consider $\theta_2 = 0, ~\theta_1 = \pi /2$,
the probabilities of $ | 0\rangle $ and $ | 1\rangle$
will remain $\frac{1}{2}$ even after the
unitary transformation has been applied.

This may also be seen from the point of view of information.
A computation is a mapping from an initial sequence to
the solution sequence, where both these sequences may be
considered to be binary. In classical computing, small noise 
added to 
the initial sequence  bits is filtered out using techniques
of discretization.
But in quantum computing, we face the
impossibility of distinguishing between amplitudes with
the multiplier $e^{i \theta}$.

If the quantum register cannot be properly initialized, the
algorithms will not work as desired.

\subsection*{Error correction}

A realistic model of computing must address the
problem of random errors.
In the circuit model, small errors would creep in
state preparation and in the implementation of the
gate operations that constitute the unitary 
transformations.

Error correction, intuitively and in classical theory,
implies that if
\[ y = x + n,\]
where $x$ is the discrete codeword, $n$ is analog noise, and
$y$ is the analog noisy codeword, one can
recover $x$  {\it completely and fully} so long as the
analog noise function $n$ is less than a certain
threshold.
If it exceeds this threshold, then also there is full
correction so long this does not happen more than
a certain number of times (the
Hamming distance for which the code is designed) at the
places the analog signal $y$ is sampled.

The hallmark of
classical  error-correcting codes is {\it the correction of
all possible small analog errors} and many others which
exceed the thresholds associated with the code alphabet.
This full correction of all possible small analog errors is
beyond the capability of the proposed quantum error-correcting codes.

This definition of error correction in classical theory
is not merely a matter of convention.
In the communication process the errors are
analog and, therefore, all possible small errors
must be corrected by error-correcting codes.
To someone who looks at classical error-correction
theory as an outsider, it may
appear that one only needs to fix bit flips.
In reality, small analog errors, occurring on
all the bits, are first removed by the use
of clamping and hard-limiting.

Since the definition of a qubit includes arbitrary phase,
it is necessary to consider errors from the perspective
of the quantum state and not just from that of final measurement.
Just as in the classical theory it is implicitly accepted
that all possible small analog errors have already been
corrected by means of an appropriate  thresholding operation,
we must define correction of small analog
phase errors as a requirement for quantum error correction.
This is something that the proposed quantum error correction
schemes are unable to do [12].

\subsection*{Statistics} 
Classical particles are distinct whereas quantum particles
are indistinguishable if they are part of the same quantum state.
Thus it becomes impossible for us to distinguish between
01 and 10 or between 001, 010, and 100, {\it before} the measurement
is made.
But the circuit
model
considers each particle to carry unique information, albeit
in a superposition.

The model
 does not consider boson/fermion statistics [13] which prevent the identification
of a qubit with any specific atom or particle within the system.
This, in turn, should make
it impossible to distinguish between
the different wires of the circuit, but in the model each wire
is uniquely labeled.

\subsection*{Hierarchy of computation models}
There may be a hierarchy of models of varying 
computational power that lie between classical and
quantum paradigms.

We know that the quantum circuit model and others that
are equivalent to it have computational power greater
than that of classical computers. But can we find other
models, still not fully quantum, that will be even more
powerful?
Knill and Laflamme have argued that if the initial state
were highly mixed one could under certain
conditions obtain
efficient solutions to some problems compared to classical
techniques [14].
This suggests that a hierarchy might very well exist.

Imposing further constraints such as
indistinguishability of the particles
may lead to computing power less than that of
the quantum circuit model.
This question should be of interest
to computer science theorists.

\subsection*{Do useful quantum computing models exist?}

Although the common quantum circuit model is not realistic,
we should not be pessimistic about the plan to devise
quantum computers.
Physical processes in the microworld unfold according to quantum
mechanics and this is enough for us 
to seek a paradigm for
computation that satisfies all the rules of
quantum mechanics. 
One would expect that in this paradigm some
problems will be solved
faster than by the fastest classical computer by virtue
of the parallelism
of quantum states.

For example, it is believed that the protein-folding problem
is NP-complete [15], yet nature performs the folding in 
a second or so, and it is plausible that this is due to the
quantum basis of the underlying chemical process.
Furthermore, the use of 
quantum apparatus offers  
an exponential edge over classical apparatus [16], 
providing us with more assurance 
that useful models of quantum computing do exist.

A realistic model of quantum computing must 
ensure that the questions of preparation of
pure states and that of boson/fermion statistics
for a quantum state
are not ignored.
It would also require a realistic method of error-correction.

\subsection*{Conclusion}

There remains the question of practical implementation
of the circuit framework without even
considering the issues raised in this note. The
requirements are so stringent
so as to make the computer physically unrealizable [17].

As
a mathematical construct,
the idea of the quantum computer will continue to provide useful insights
into the nature of the information process.

\subsection*{References}

\begin{enumerate}

\item 
See
A.M. Steane, ``Quantum computing,''
{\it Rept.Prog.Phys.} 61, 117 (1998),
quant-ph/9708022, for an account of
cryptography,
dense coding and teleportation;
S. Kak, 
``Representation of entangled states,''
quant-ph/0302118.

\item
For example, see P. Shor, ``Introduction to
quantum algorithms.'' quant-ph/0005003. Shor
acknowledges the unrealizability of the model on page 3.

\item
D. Deutsch,
``Quantum theory, the Church-Turing principle and the 
universal quantum computer.''
{\it Proc. Roy. Soc. London Ser. A,} {\bf 400,} 97-117 (1985).

\item
A. Peres, {\it Quantum Theory: Concepts and Methods}.
Kluwer Academic, Dordrecht, 1995, page 24.

\item
E. Fredkin and T. Toffoli, ``Conservative logic.''
{\it Int. J. Theor. Phys.,} {\bf 21,} 219-253 (1982).

\item
R. Landauer, ``Information is physical.''
 {\it Phys. Today,} {\bf 44,} 23-29 (1991);
R. Landauer, ``The physical nature of information.''
{\it Phys. Lett. A,} {\bf 217,} 188 (1996).

\item
E. Fredkin and T. Toffoli, {\it op cit.}

\item
A. Peres, {\it op cit.}, page 30.

\item
R. Laflamme {\it et al}, ``Introduction 
to NMR quantum information processing.''
quant-ph/0207172.

\item
S. Kak, ``The initialization problem in quantum computing.''
{\it Foundations of Physics,} {\bf 29,} 267-279 (1999); 
quant-ph/9805002.
\item
S. Kak, ``Rotating a qubit.''
{\it Information Sciences,} {\bf 128,} 149-154 (2000); quant-ph/9910107.

\item
S. Kak, ``General qubit errors cannot be corrected.''
{\it Information Sciences,} {\bf 152,} 195-202 (2003); 
quant-ph/0206144.

\item
S. Kak, ``Statistical constraints on state preparation for a quantum 
computer.''
{\it Pramana,} {\bf 57,} 683-688 (2001) ; quant-ph/0010109.


\item
E. Knill and R. Laflamme, ``On the power of one bit of quantum information.''
{\it Phys. Rev. Lett.,}
{\bf 81,} 5672 (1998).

\item
A. Fraenkel, ``Complexity of protein folding.''
     {\it Bulletin of Mathematical Biology,} {\bf 55,} 1199 (1993),\\ 
R. Unger and J. Moult, 
``Finding the lowest free energy conformation of a protein is an NP-hard problem: proof and implications.'' 
     {\it Bulletin of Mathematical Biology,} {\bf 55,} 1183-1198, 1993.\\
B. Berger and T. Leighton, 
``Protein folding in the hydrophilic-hydrophobic (HP) model is NP-complete.'' 
     {\it Journal of Computational Biology,} {\bf 5}, 27 (1998).


\item
S. Kak, ``Speed of computation and simulation.''
{\it Foundations of Physics,} {\bf 26,} 1375-1386 (1996); quant-ph/9804047.

\item
M.I. Dyakonov, ``Quantum computing: a view from the
enemy camp.'' cond-ph/0110326.

\end{enumerate}
 
\end{document}